\documentclass{iopconfser}
\usepackage{graphicx}
\usepackage{amsmath}
\usepackage{cite}
\usepackage{hyperref}
\providecommand{\keywords}[1]
{
  \textbf{\textit{Keywords:  }} #1
} 

\begin{document}

\title{Numerical search for three-body periodic free-fall orbits with central symmetry}

\author{I. Hristov$^{1}$, R. Hristova$^{1}$, T. Puzynina$^{2}$, Z. Sharipov$^{2}$, Z. Tukhliev$^{2}$}

\affil{$^{1}$ Sofia University ``St. Kliment Ohridski'', Faculty of Mathematics and Informatics, James Bourchier blvd. 5,
1136 Sofia, Bulgaria}
\vspace{0.3 cm}
\affil{$^{2}$ 
Joint Institute for Nuclear Research, Meshcheryakov Laboratory of Information Technologies, Joliot-Curie 6,
141980 Dubna, Moscow region, Russia}

\email{ivanh@fmi.uni-sofia.bg, radoslava@fmi.uni-sofia.bg, zarif@jinr.ru} 

\vspace{1.0 cm}
\begin{flushright} 
{\it{To the bright memory of our teacher Igor Viktorovich Puzynin}}
\end{flushright}

\begin{abstract}
A specialized high-precision numerical search for equal-mass collisionless three-body periodic free-fall orbits with central symmetry is conducted.
The search is based on Newton's method with initial approximations obtained by the grid-search method.
Instead of solving the standard periodicity equation on the entire period a quarter-period equation 
that also characterizes the periodic orbits is solved.
The number of the known orbits from the class is significantly enlarged.
The linear stability of the orbits is also investigated. All of them are unstable.
A discussion in relation to the efficiency of Newton's method applied with grid-search initial approximations is held.
\end{abstract}

{\keywords{three-body problem, collisionless periodic free-fall orbits, central symmetry, linear stability, Lyapunov exponents, grid-search method, Newton's method}}

\section{Introduction}
\label{s:Introduction}

This work is motivated by a recent paper of Richard Montgomery \cite{Dropping_bodies:2023}, where a class of equal-mass collisionless three-body periodic free-fall orbits with central symmetry is analyzed. The orbits were previously found in \cite{Li:2019}. From the analysis in \cite{Dropping_bodies:2023} follows that this class of orbits has a quarter-period property. More specifically, at  each quarter of the period, free-fall initial conditions \cite{Dropping_bodies:2023} alternate with initial conditions with isosceles collinear configuration \cite{Suvakov:2013, Suvakov:2014, Li:2017, Li:2018, Hristovi:2024}. 
To conduct a specialized (targeted) search for these special orbits in the present paper we use exactly this quarter-period property.
In fact, the considered class is the intersection of the two larger classes of periodic orbits: the class of periodic orbits that at some instant have free-fall  initial conditions and  the class of periodic orbits that at some instant have initial conditions with isosceles collinear configuration. 
Particularly the orbits can be obtained as a result of searching for periodic orbits in either of the two larger classes  and actually this is the way they were obtained for the first time. Let us agree that from this point onwards the term ``Euler half-twist initial conditions'' used in \cite{Hristovi:2024} will be used in place of the term ``Initial conditions with isosceles collinear configuration'' from \cite{Suvakov:2013, Suvakov:2014, Li:2017, Li:2018}. Furthermore, the term ``initial conditions'' will be used more freely, not only for time $t=0$, but also for other instants.

 We already have two examples of efficient numerical searches (with a large number of new periodic orbits) based on half-period properties of the solutions. The two examples are just  for free-fall orbits in \cite{free_fall:2024, free_fall_stab:2024} and Euler half-twist initial conditions' orbits in \cite{Hristovi:2024}. In the first case the property that the solutions again have free-fall initial conditions at the half period is used. In the second case the property that the solutions again have half-twist initial conditions at the half period is used. 
The efficient numerical searches in these papers are  based, like those in \cite{Li:2019, Li:2017, Li:2018}, on the grid-search method in combination with Newton's method. The candidates for periodic solutions (the initial approximations for Newton's method) are computed  by the grid-search method applied on the initial conditions' search domain. So, the main framework of the searches in \cite{free_fall:2024, free_fall_stab:2024, Hristovi:2024} is the same as for the previous ones. The difference of the approach in \cite{free_fall:2024, free_fall_stab:2024, Hristovi:2024} from that in \cite{Li:2019, Li:2017, Li:2018} is that instead of solving the standard periodicity equation on the entire period, equations which also characterize the periodic conditions, but on a half period, are solved. Namely, the equations that express the mentioned above half-period properties are solved. 

 In this work we go further in reducing the considered part of the period $T$. We make a specialized search for the class of orbits in \cite{Dropping_bodies:2023} as follows: Instead of solving the standard periodicity equation on the entire period, another equation, which also characterizes the periodic solutions, but on a quarter period, is solved. The equation that is solved expresses the conditions that the periodic solutions have Euler half-twist conditions at $t=0$ and free-fall initial conditions at $t=T/4$. Let us mention that  this search can also be done with the exchange of the two types of initial conditions: free-fall initial conditions at $t=0$ and Euler half-twist initial conditions at $t=T/4$. 
 Actually, the considered quarter-period approach is somehow like the class itself - the ``intersection'' of the half-period approaches of the larger classes. In this sense the present work can be regarded as a continuation of works \cite{free_fall:2024, free_fall_stab:2024, Hristovi:2024}.

 Using the new quarter-period approach we succeed to expand significantly the database of the so far known solutions of this interesting (at least from a purely mathematical point of view) class of periodic orbits. Our goal, however,  is not only to present a large number of new periodic orbits. In this example, we also want to demonstrate how important the consideration of equations that characterize the periodic
 solutions from a given class only on a part of the period is for the efficiency of searching for periodic orbits for a
 chaotic system (like the three-body system). Of course, the ability to examine a smaller part of the period comes at the cost of performing a specialized search on a narrower class of orbits (usually orbits with additional symmetries). On the other hand, considering a smaller part of the period serves as a good opportunity to test and analyse the efficiency of the grid-search method in combination with Newton's method.  As we will explain in terms of Lyapunov exponents, the smaller the part of the period, the better the results. The results obtained here allow us to conclude that although we are learning how to make our searches  better and better, they still do not seem to be that close to exhaustive searches for limited periods (at least in the chaotic regions of the search domains).
 
 Actually, this series of works  was initially inspired by the ideas in \cite{Suvakov:2014b, Suvakov:2016}, where a third-period property of special solutions called choreographies was used for a specialized numerical search. This property was used in our recent paper \cite{Grid:2023}, where we solve an equation on a third of the period.
 
 The paper consists of 6 sections. After the Introduction, in Section \ref{s:Mathematical model} we present the differential equations describing the motion of three bodies and the two types of considered initial conditions: Euler half-twist initial conditions and free-fall initial conditions.
Section \ref{s:Description of the class} describes the symmetries of the considered orbits.
Section \ref{s:Numerical methods} explains the numerical methods used.
The numerical results and discussion of the efficiency of Newton's method are presented in Section \ref{s:Numerical results}. In Section \ref{s:Conclusions} we draw the conclusions.

\section{Mathematical model}
\label{s:Mathematical model}

\subsection{Differential equations}
\label{ss:Differential equations}

The motion of three bodies in space interacting each other by gravitational forces is described by the following second order ODEs system:
\begin{equation}
\label{l:equation1}
m_i\ddot{r}_i=\sum_{j=1,j\neq i}^{3}G m_i m_j \frac{(r_j-r_i)}{{\|r_j -r_i\|}^3}, i=1,2,3.
\end{equation}
In system (\ref{l:equation1}) $r_i$ is the vector of positions of the $i$-th body, $m_i$ is the mass of the $i$-th body, 
$G$ is the gravitational constant. 
As we consider zero angular momentum solutions, the motion is planar.
Hence, the vectors $r_i$, $\dot{r}_i$ in our case have two components - $r_i=(x_i, y_i)$,
$\dot{r}_i=(\dot{x}_i, \dot{y}_i).$
Equal masses $m_1=m_2=m_3$ and normalization with $G=1$ and $m_i=1$ are considered in this work.
If the dependent variables ${vx}_i$ and ${vy}_i$ are introduced, so that ${vx}_i=\dot{x}_i, {vy}_i=\dot{y}_i,$
then the second order system (\ref{l:equation1}) can be written as a first order one:
\begin{equation}
\label{l:equation2}
\dot{x}_i={vx}_i, \hspace{0.1 cm} \dot{y}_i={vy}_i,  \hspace{0.1 cm} \dot{vx}_i=\sum_{j=1,j\neq i}^{3}\frac{(x_j-x_i)}{{\|r_j -r_i\|}^3},  \hspace{0.1 cm} \dot{vy}_i=\sum_{j=1,j\neq i}^{3}\frac{(y_j-y_i)}{{\|r_j -r_i\|}^3}, \hspace{0.1 cm} i=1,2,3.
\end{equation}
In our numerical study we use this more convenient first order form.
We have the following 12-component unknown vector $X(t)$ of positions and velocities:
$$X(t)={(x_1(t), y_1(t), x_2(t), y_2(t), x_3(t), y_3(t), {vx}_1(t), {vy}_1(t), {vx}_2(t), {vy}_2(t), {vx}_3(t), {vy}_3(t))}^\top$$

\subsection{Euler half-twist initial conditions}
\label{ss:Euler initial conditions}

\begin{figure}
\centerline{\includegraphics[width=0.4\columnwidth,,keepaspectratio]{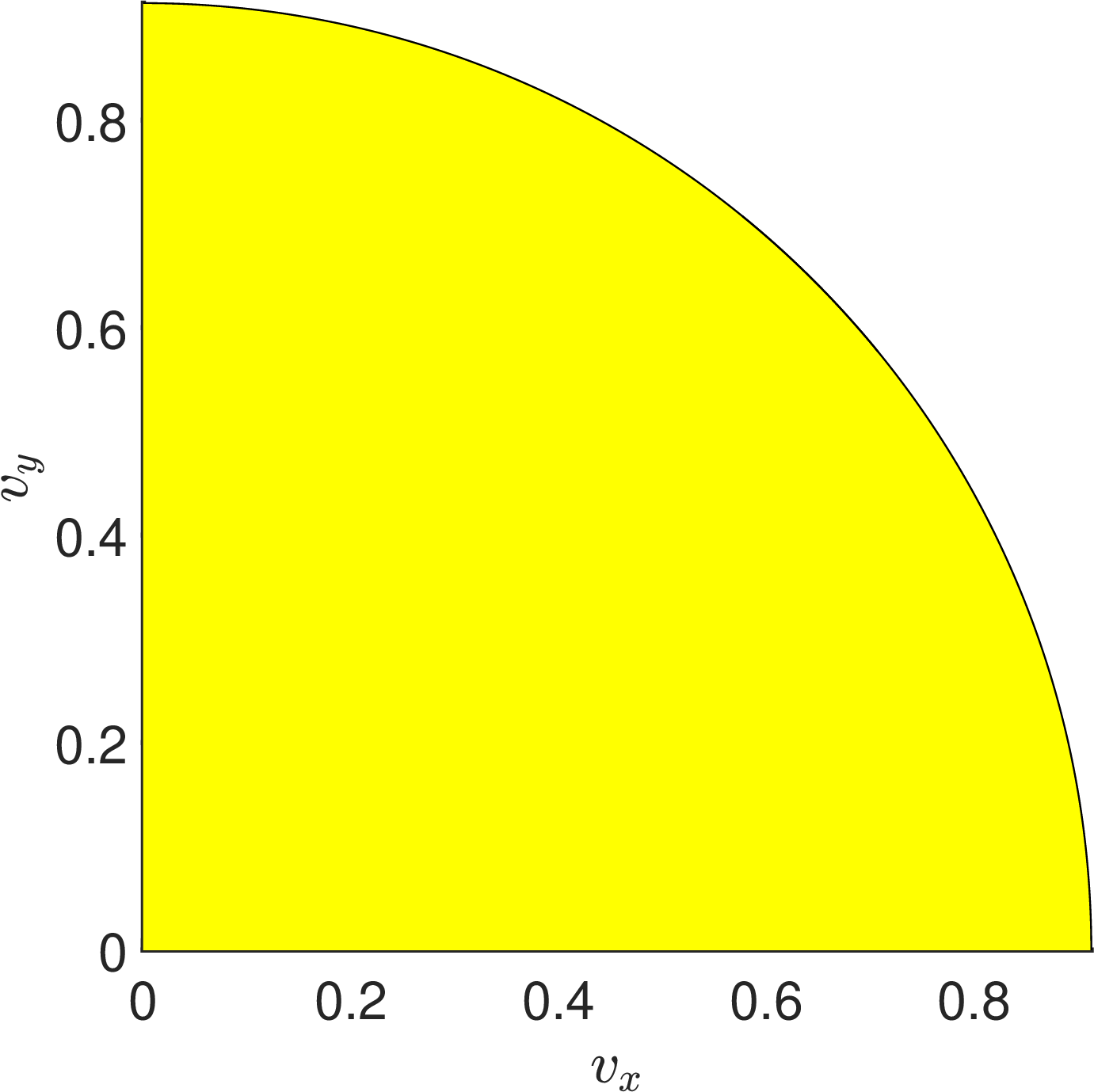}}
\caption{Euler half-twist initial conditions' domain}
\label{f:Edomain}
\end{figure}

Euler half-twist initial conditions (i.c.) with two parameters $v_x>0$, $v_y>0$ \cite{Suvakov:2013, Suvakov:2014, Li:2017, Hristovi:2024} read as follows:
\begin{center}
\begin{equation}
\label{initcond}
\begin{split}
(x_1(0),y_1(0))=(-1,0), \hspace{0.2 cm} (x_2(0),y_2(0))=(1,0), \hspace{0.2 cm} (x_3(0),y_3(0))=(0,0) \hspace{0.5 cm} \\
({vx}_1(0),{vy}_1(0))=({vx}_2(0),{vy}_2(0))=(v_x,v_y)  \hspace{2.5 cm}\\
({vx}_3(0),{vy}_3(0))=-2({vx}_1(0),{vy}_1(0))=(-2v_x, -2v_y) \hspace{1.8 cm}
\end{split}
\end{equation}
\end{center}
For initial positions fixed as above, the $2D$ space of velocities described
by equation (\ref{initcond}) is precisely the space of  all  velocities for which
the linear and angular momentum are zero and  for which the moment of inertia $I = {x_1}^2 + {y_1}^2 +  {x_2}^2 + {y_2}^2 + {x_3}^2 + {y_3}^2$
has an extremum at $t=0$ \cite{Hristovi:2024}.
The system has energy $E(v_x, v_y)=-2.5+3({v_x}^2+ {v_y}^2)$.
As negative energies are necessary for bounded motions (what the periodic motions are) \cite{Suvakov:2014},
the $2D$ search domain is those bounded  by $v_x=0$ and $v_y=0$ axis and $E=0$ curve in Figure \ref{f:Edomain}. 
The circular curve defining the points with zero energy (the $E=0$ curve) in the quadrant I of the $(v_x,v_y)-$plane is defined by the function: $v_y=\sqrt{5/6-{v_x}^2}, v_x\in[0,\sqrt{5/6}], \sqrt{5/6}\approx 0.91287$.  If we denote the solution of equation (\ref{l:equation2}) with initial conditions 
(\ref{initcond}) with $X(v_x,v_y,t)$ and look for periodic orbits, then the goal is to find triplets $(v_x,v_y,T)$ for which the periodicity condition
(nonlinear equation)
$$X(v_x,v_y,T)=X(v_x,v_y,0)$$
is satisfied.

\subsection{Free-fall initial conditions}
\label{ss:Free-fall initial conditions}

\begin{figure}
\centerline{\includegraphics[width=0.4\columnwidth,,keepaspectratio]{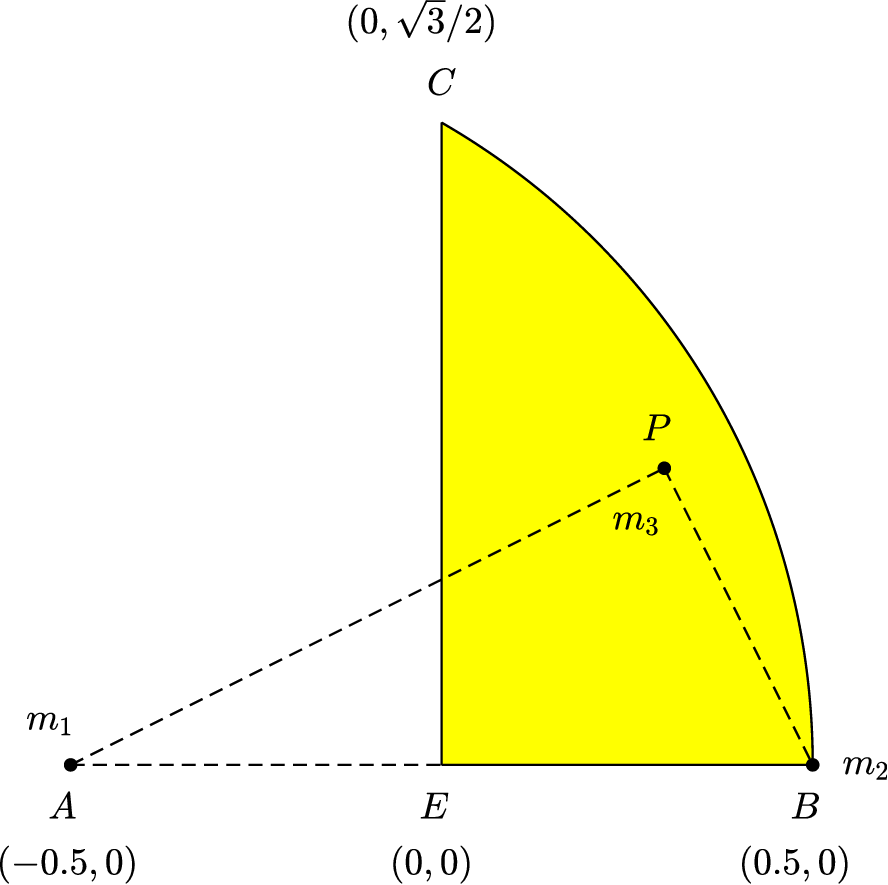}}
\caption{Agekyan-Anosova's domain (free-fall initial conditions' domain)}
\label{f:AAdomain}
\end{figure}

Free-fall initial conditions (i.c.) mean that bodies are with zero velocities (they are stopped) at $t=0$.
Of course, for these initial conditions we again have zero linear and zero angular momentum.
All different shapes of the initial triangle at $t=0$ can be generated in the following way (see Figure \ref{f:AAdomain}).
Body $1$ and body $2$ are fixed at points $A=(-0.5, 0)$ and $B=(0.5,0)$ respectively and then 
body $3$ is placed at all possible positions of point $P(p_x, p_y)$ in the domain bounded by the $x$ and $y$ axes and the
arc of a circle with a center $A$ and a radius of 1. This is the $2D$ free-fall initial conditions' domain called the Agekyan-Anosova's domain (see \cite{Agekyan_Anosova:1967} for details). Free-fall i.c. with two parameters $p_x>0$, $p_y>0$ (the coordinates of point $P$) read as follows:
\begin{center}
\begin{equation}
\label{initcond2}
\begin{split}
(x_1(0),y_1(0))=(-0.5,0), \hspace{0.2 cm} (x_2(0),y_2(0))=(0.5,0), \hspace{0.2 cm} (x_3(0),y_3(0))=(p_x,p_y)\\
({vx}_1(0),{vy}_1(0))=({vx}_2(0),{vy}_2(0))=({vx}_3(0),{vy}_3(0))=(0,0)\\
\end{split}
\end{equation}
\end{center}
The energy in the case of free-fall i.c. is: 
$$E(p_x, p_y)=-1-\frac{1}{\sqrt{(p_x+0.5)^2+p_y^2}}-\frac{1}{\sqrt{(p_x-0.5)^2+p_y^2}}$$ 
and is always negative.
If we denote the solution of equation (\ref{l:equation2}) with initial conditions 
(\ref{initcond2}) with $X(p_x,p_y,t)$ and look for periodic orbits, then the goal is to find triplets $(p_x,p_y,T)$ for which the periodicity condition
(nonlinear equation):
$$X(p_x,p_y,T)=X(p_x,p_y,0)$$
is satisfied. 

\section{Description of the symmetries of the periodic orbits}
\label{s:Description of the class}
\begin{figure}
\centerline{\includegraphics[scale=0.52]{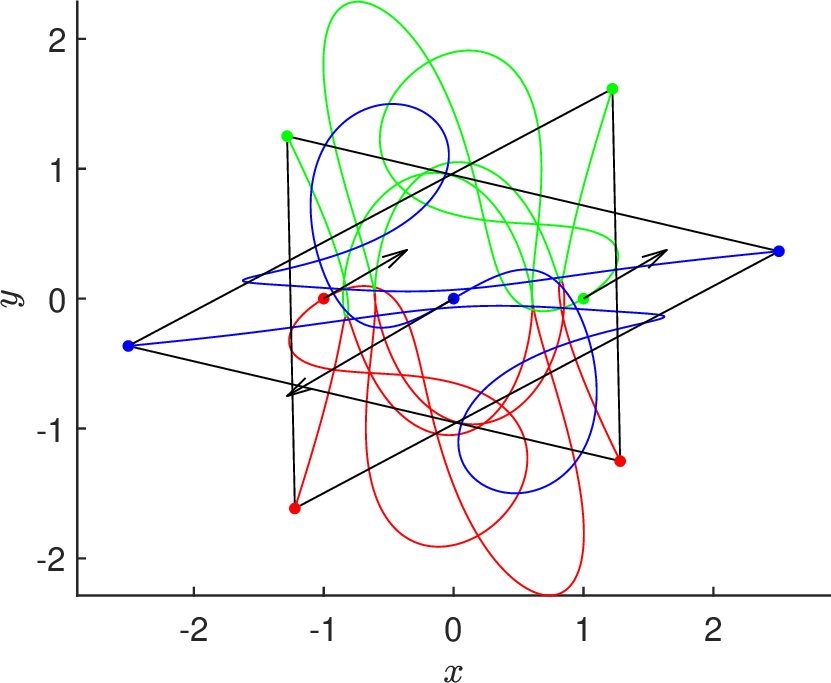}
\includegraphics[scale=0.52]{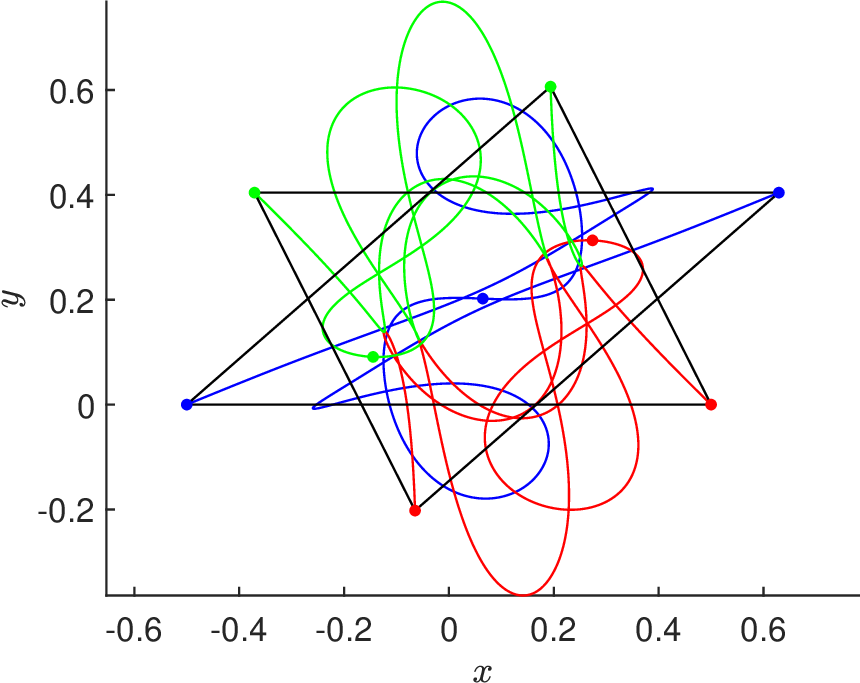}}
\vspace{0.5 cm}
\caption{{Real space plots for $F_{14}(1,1,1)$ from \cite{Li:2019}: Left -- Euler half-twist i.c.;  Right -- free-fall i.c.
The ``stop'' triangles are shown. The positions of the bodies at $t=0,T/4,T/2,3T/4$ are given with circles.}}
\label{fig:F14}
\end{figure}

We will describe the symmetries using as an example the solution $F_{14}(1,1,1)$ found in \cite{Li:2019} (see Figure \ref{fig:F14}).
The same example is used in \cite{Dropping_bodies:2023}.
Real space plot of the solution in the Euler half-twist initial conditions' domain is given in the left plot. Initial velocities are presented with arrows.
Real space plot of the solution in the free-fall initial conditions' domain is given in the right plot.
Note that the same periodic orbit is presented by solutions with different energies in the two domains and hence 
the left and right plots in Figure \ref{fig:F14} are in different scales. 
So, to obtain identical solutions they need to be scaled each other according to the rules from \cite{Suvakov:2014}:
If $r\rightarrow \alpha r$, where $\alpha$ is the scaling factor, then $t \rightarrow \alpha^{\frac{3}{2}} t$, $v \rightarrow v/\sqrt{\alpha}$, $E \rightarrow \alpha^{-1}E$. Note also that some isometry transformation in the plane and consistent mass labeling are needed 
to match the curves in the left and right plots.

As we have free-fall periodic solutions the bodies shuttle back and forth between two ``stop'' triangles \cite{Dropping_bodies:2023}.  For the considered class of solutions the ``stop'' triangles are additionally congruent (as unlabeled ones) and $180^{\circ}$ rotated  around the center of the mass \cite{Dropping_bodies:2023} (equivalent to having a central inversion with center the center of the mass). The  ``stop'' triangles are congruent as unlabeled ones, because two of the masses (the green and red bodies in Figure \ref{fig:F14}) are swapped (interchanged) in the congruent triangles (see ``Symmetry puzzles'' section in  \cite{Dropping_bodies:2023}). If we assume such a $180^{\circ}$-rotation with two swapped masses, then the entire central symmetry of the curves in the real space (taking into account the swapped masses) follows. It also follows that if we have free-fall i.c. at $t=0$,  then  at $t=T/4$ and $t=3T/4$ the solution has Euler half-twist i.c. and the body which is not swapped (the blue body in Figure \ref{fig:F14}) is in the center of the mass at these instants.
At $t=T/4$ and $t=3T/4$ the bodies have the same positions, but the velocities are opposite. Overall, we can describe the motion of the bodies as an alternation of  Euler half-twist i.c. and free-fall i.c. at each quarter of the period. The data with 35 correct digits for solution $F_{14}(1,1,1)$ from \cite{Li:2019} is given in Tables 1 and 2. All given digits in \cite{Li:2019} for the free-fall i.c. coincide with our first digits. Note also that the scale-invariant periods $T^{*}=T|E|^{3/2}$ in Tables 1 and 2 need to be, and they are the same, as the solution is the same.

There are four ways to conduct a specialized search,  as these periodic orbits can be characterized equivalently by four different  properties.
The first way is to search for orbits that have free-fall i.c. at $t=0$ and the  ``stop'' triangle at $t=T/2$ is obtained by a $180^{\circ}$ rotation of the initial one. The second way is to search for orbits that  have Euler half-twist i.c. at $t=0$ and again Euler half-twist i.c. 
at $t=T/2$ with the same positions and opposite velocities. The third way is to search for orbits that have free-fall i.c. at $t=0$ and Euler half-twist i.c. at $t=T/4$. The fourth way is to search for orbits with Euler half-twist i.c. at $t=0$ and free-fall i.c. at $t=T/4$. The first two ways use half-period properties and (as we will see) are not quite so good (not that efficient). In this work we use the fourth way.

It is of interest to note that although the equal-mass free-fall solutions from \cite{Li:2019} obtained by searching for free-fall orbits are analyzed in \cite{Dropping_bodies:2023}, this kind of orbits were obtained a little earlier by Li, Jing and Liao \cite{Li:2018} as a result of an Euler-half twist initial conditions' search for unequal mass periodic orbits. Such an orbit is for example the orbit $II.D_2^{i.c}(2)$ in section 2 ``Periodic orbits with unequal mass'' on the web-site in  \cite{Website:Liao}. The explanation to this result is the existence of a generalization of the considered class of orbits when the different mass body for  the Euler half-twist i.c. is in the middle between the two other ones with equal masses. The swapped masses in the rotated ``stop'' triangles are exactly the equal masses.
\begin{table}[h]
 \begin{tabular}{ p{7cm} p{7cm}}
 \hline
  $\hspace{3cm}v_x$ & $\hspace{3cm}v_y$\\
  $\hspace{3cm}T$ & $\hspace{3cm}T^*$\\
  \hline
  0.6414776117623788428234377297741194e0 & 0.37534705681712968770351929225673456e0 \\
  0.66315883746492982214861887011039939e2 & 0.51316046066182872741916938377446188e2 \\
 \hline
\end{tabular}
{\caption {35 correct digits for the Euler half-twist i.c. of solution $F_{14}(1,1,1)$ from \cite{Li:2019}. }}
\label{tab:table1}
\end{table}
\begin{table}[h]
 \begin{tabular}{ p{7cm} p{7cm}}
 \hline
  $\hspace{3cm}p_x$ & $\hspace{3cm}p_y$\\
  $\hspace{3cm}T$ & $\hspace{3cm}T^*$\\
  \hline
  0.19382337571695823384109237522588605e0 & 0.60637511021612394148276815543012096e0 \\
  0.76482262218471020895008233552139143e1 & 0.51316046066182872741916938377446188e2 \\
 \hline
\end{tabular}
{\caption {35 correct digits for the free-fall i.c. of solution $F_{14}(1,1,1)$ from \cite{Li:2019}. }}
\label{tab:table2}
\end{table}

\section{Numerical methods}
\label{s:Numerical methods}
To better understand the idea of the quarter-period approach, we need to show how the linear systems from Newton's method look. All other details are sketched and references for them are given. Let us agree to use the same notation for the triplet $(v_x,v_y,T)$ and its approximation.
\subsection{The linear system at each Newton's iteration for the periodicity equation}
\label{ss:Periodicity equation}
First we will explain how the standard periodicity equation for the Euler half-twist i.c.
$$X(v_x,v_y,T)=X(v_x,v_y,0)$$
is solved. After all, we always end up solving this equation.
The return proximity function is defined as follows \cite{Hristovi:2024}:
$$
R(t)=\|X(v_x,v_y,t) - X(v_x,v_y,0)\|_{2}, \quad  t>0
$$
The triplet $(v_x,v_y,T)$  corresponds to a periodic solution with period $T>0$ only if $R(T)=0$.
For an approximate solution with approximations $(v_x,v_y,T)$, $R(T)$ measures how well the periodicity equation is satisfied 
(how close to a periodic solution we are).
$R(T)$ is the residual in terms of which we define the convergence of Newton's method for the periodicity equation.

Let $(v_x,v_v, T)$ be an approximation for a periodic orbit, i.e. $X(v_x,v_y,T)\approx X(v_x,v_y,0).$
With Newton's method these approximations are improved with corrections $\Delta v_x,
\Delta v_y, \Delta T$ by expanding the periodicity equation:\\
$$X(v_x+\Delta v_x,v_y+\Delta v_y,T+\Delta T)= X(v_x+\Delta v_x,v_y+\Delta v_y,0)$$
in a multivariable linear approximation \cite{Hristovi:2024}: 

\begin{equation}
\label{l:Newton1}
\begin{pmatrix}
x_1(T) \\
y_1(T) \\
x_2(T) \\
y_2(T) \\
x_3(T) \\
y_3(T) \\
{vx}_1(T)\\
{vy}_1(T)\\
{vx}_2(T)\\
{vy}_2(T)\\
{vx}_3(T)\\
{vy}_3(T)
\end{pmatrix}
+
\begin{pmatrix}
\frac{\partial x_1}{\partial v_x}(T) & \frac{\partial x_1}{\partial v_y}(T) & \dot{x}_1(T)\\
\frac{\partial y_1}{\partial v_x}(T) & \frac{\partial y_1}{\partial v_y}(T) & \dot{y}_1(T)\\
\frac{\partial x_2}{\partial v_x}(T) & \frac{\partial x_2}{\partial v_y}(T) & \dot{x}_2(T)\\
\frac{\partial y_2}{\partial v_x}(T) & \frac{\partial y_2}{\partial v_y}(T) & \dot{y}_2(T)\\
\frac{\partial x_3}{\partial v_x}(T) & \frac{\partial x_3}{\partial v_y}(T) & \dot{x}_3(T)\\
\frac{\partial y_3}{\partial v_x}(T) & \frac{\partial y_3}{\partial v_y}(T) & \dot{y}_3(T)\\
\frac{\partial {vx}_1}{\partial v_x}(T) & \frac{\partial {vx}_1}{\partial v_y}(T) & \dot{vx}_1(T)\\
\frac{\partial {vy}_1}{\partial v_x}(T) & \frac{\partial {vy}_1}{\partial v_y}(T) & \dot{vy}_1(T)\\
\frac{\partial {vx}_2}{\partial v_x}(T) & \frac{\partial {vx}_2}{\partial v_y}(T) & \dot{vx}_2(T)\\
\frac{\partial {vy}_2}{\partial v_x}(T) & \frac{\partial {vy}_2}{\partial v_y}(T) & \dot{vy}_2(T)\\
\frac{\partial {vx}_3}{\partial v_x}(T) & \frac{\partial {vx}_3}{\partial v_y}(T) & \dot{vx}_3(T)\\
\frac{\partial {vy}_3}{\partial v_x}(T) & \frac{\partial {vy}_3}{\partial v_y}(T) & \dot{vy}_3(T)
\end{pmatrix}
\begin{pmatrix}
\Delta v_x \\
\Delta v_y \\
\Delta T
\end{pmatrix}
=
\begin{pmatrix}
x_1(0) \\
y_1(0)  \\
x_2(0) \\
y_2(0) \\
x_3(0)  \\
y_3(0)  \\
{vx}_1(0) + \Delta v_x\\
{vy}_1(0) + \Delta v_y\\
{vx}_2(0) + \Delta v_x\\
{vy}_2(0) + \Delta v_y\\
{vx}_3(0)- 2 \Delta v_x\\
{vy}_3(0)- 2 \Delta v_y
\end{pmatrix}
\end{equation}
Equation (\ref{l:Newton1}) is solved with respect to the corrections $\Delta v_x, \Delta v_y, \Delta T$ 
at each Newton's iteration when solving the periodicity equation for the Euler half-twist initial conditions' domain.

\subsection{The linear system at each Newton's iteration for the quarter-period equation}
Let us recall that our approach is to search for orbits with Euler half-twist i.c. at $t=0$ and free-fall i.c. at $t=T/4$.
If we denote the second half (velocities' half) part of $X(v_x,v_y,t)$ with $V(v_x,v_y,t)$:
$$V(v_x,v_y,t)={({vx}_1(t), {vy}_1(t), {vx}_2(t), {vy}_2(t), {vx}_3(t), {vy}_3(t))}^\top$$
the quarter-period equation becomes: 
$$V(v_x,v_y,T/4)=0$$
The proximity function for this equation is:
$$R_{vel}(t)=\|V(v_x, v_y, t)\|_{2}, \hspace{0.2 cm} t>0$$
Let $\overline{T}$ be an approximation of $T/4$.
For an approximate solution with approximations $(v_x,v_y,\overline{T})$, $R_{vel}(\overline{T})$ measures how well we satisfy the quarter-period equation.
Let $(v_x,v_v,\overline{T})$ be an approximation for a periodic orbit, i.e. 
$V(v_x,v_y,\overline{T})\approx 0.$
These approximations are improved with corrections $\Delta v_x,
\Delta v_y, \Delta \overline{T}$ by expanding the equation:
$$V(v_x+\Delta v_x, v_y+\Delta v_y, \overline{T}+ \Delta \overline{T})=0$$
in a multivariable linear approximation:

\begin{equation}
\label{l:Newton2}
\begin{pmatrix}
{vx}_1(\overline{T})\\
{vy}_1(\overline{T})\\
{vx}_2(\overline{T})\\
{vy}_2(\overline{T})\\
{vx}_3(\overline{T})\\
{vy}_3(\overline{T})
\end{pmatrix}
+
\begin{pmatrix}

\frac{\partial {vx}_1}{\partial v_x}(\overline{T}) & \frac{\partial {vx}_1}{\partial v_y}(\overline{T}) & \dot{vx}_1(\overline{T})\\
\frac{\partial {vy}_1}{\partial v_x}(\overline{T}) & \frac{\partial {vy}_1}{\partial v_y}(\overline{T}) & \dot{vy}_1(\overline{T})\\
\frac{\partial {vx}_2}{\partial v_x}(\overline{T}) & \frac{\partial {vx}_2}{\partial v_y}(\overline{T}) & \dot{vx}_2(\overline{T})\\
\frac{\partial {vy}_2}{\partial v_x}(\overline{T}) & \frac{\partial {vy}_2}{\partial v_y}(\overline{T}) & \dot{vy}_2(\overline{T})\\
\frac{\partial {vx}_3}{\partial v_x}(\overline{T}) & \frac{\partial {vx}_3}{\partial v_y}(\overline{T}) & \dot{vx}_3(\overline{T})\\
\frac{\partial {vy}_3}{\partial v_x}(\overline{T}) & \frac{\partial {vy}_3}{\partial v_y}(\overline{T}) & \dot{vy}_3(\overline{T})\\
\end{pmatrix}
\begin{pmatrix}
\Delta v_x \\
\Delta v_y \\
\Delta \overline{T}
\end{pmatrix}
=
\begin{pmatrix}
0\\
0\\
0\\
0\\
0\\
0\\
\end{pmatrix}
\end{equation}
Equation (\ref{l:Newton2}) is solved at each Newton's iteration when solving the quarter-period equation.
Note that if we have a solution in the Euler half-twist initial conditions' domain, then by rescaling the solution at $t=T/4$
and testing the convergence of Newton's method for the periodicity equation in the free-fall initial conditions domain
$$X(p_x,p_y,T)=X(p_x,p_y,0)$$
we easily obtain the solution in this domain. 

All linear systems for Newton's method are solved as linear least square problems using QR decompositions based
on Householder reflections \cite{Demmel}.

\subsection{Stages of the numerical search}
\label{ss:Stages of the numerical search}
The numerical search consists of three stages. At the first stage we scan the interior of the domain in Figure \ref{f:Edomain}.
A quadratic grid with a size of $3*2^{-14} \approx 1.831*10^{-4}$ corresponding to about 19 million points is introduced. At each point $(v_x,v_y)$ of the grid, the system (\ref{l:equation2}) 
with initial conditions (\ref{initcond}) is simulated up to time $T_0(v_x,v_y)/4$. In this work $T_0(v_x,v_y)=T^*_{max}/|E(v_x,v_y)|^{3/2},$ 
where $T^*_{max}=120$ is the upper limit of the considered scale-invariant periods.
The value $\overline{T}$, for which the minimum 
$$R_{vel}(\overline{T})=\min\limits_{1<t<T_0/4}R_{vel}(t)$$
is obtained, is computed. Candidates for periodic orbits become triplets $(v_x,v_v,\overline{T})$ for which $R_{vel}(\overline{T})$ is small and is a local minimum on the grid. To accurately follow the trajectories for sufficiently long times system (\ref{l:equation2}) 
with initial conditions (\ref{initcond}) is solved using the high order high precision Taylor series method 
\cite{Jorba, Barrio:2006, Barrio:2011, Izzo}. 
To compute the coefficient of the Taylor series the rules of automatic differentiation are used \cite{Jorba, Barrio:2006, Barrio:2011, Izzo}.
The order of the Taylor series method depends on the set precision.
More precisely, 22 order of Taylor series are used for each 64 bit mantissa (precision), which is an optimal choice
for the used time stepsize strategy.
44 order of Taylor series and 128 bit of precision  are used at the first stage. The order and precision are increased in the next stages of the search.
 
At the second stage we capture the periodic orbits by testing the candidates from the first stage 
for convergence with Newton's method. Newton's method is applied on the quarter-period equation with residual $R_{vel}(\overline{T})$.
At the third stage we verify the captured solution by testing the convergence of Newton's method which is now applied for the standard periodicity equation
with residual $R(T)$. The elements in the linear systems (\ref{l:Newton1}) and (\ref{l:Newton2}) from the second and third stages are computed again using
the Taylor series method by solving the system of 36 coupled differential equations for positions, velocities and their partial derivatives with respect
to $v_x$ and $v_y$. The rules of automatic differentiation for the Taylor series method are extended for the partial derivatives. 

All automatic differentiation formulas can be found in \cite{Hristov:2021}. 
More details for Newton's method, including the time stepsize strategy and the high precision verification of the periodic orbits are given in \cite{Hristovi:2024, free_fall_stab:2024}.
Our experience with Taylor series method reveals that it handles both the sensitive dependence on the initial conditions
and the close approaches of the bodies very well.

\section{Numerical results}
\label{s:Numerical results}

\subsection{ The solutions found and their linear stability}
\label{ss:The solutions found}

As a result of the specialized search, we obtained 4860 periodic orbits with $T^*<120$. 
Since the two ``stop''  triangles are congruent, the solutions are presented by one point only (by just one initial condition) in both the Euler half-twist i.c. and the free-fall i.c. domains. All found solutions are ordered by $T^*$. 

The data $(v_x, v_y, T, T^*)$  and  $(p_x, p_y, T, T^*)$ 
for the two domains is given with 100 correct digits and can be found in  \cite{Paper web-site:2025}. 
We give 100 correct digits not because they are needed to ``close'' one period well (with a small $R(T$)), but in order to show that in principle these computations can be done,  i.e. that Newton's method has regular convergence when the precision is increased. This gives us much greater confidence in the existence of these solutions and hence rigour of our results. What precision do we need anyway?
Double or extended (long double) precisions are  not sufficient to ``close'' one period well for all of the found orbits.
After all, no one expects that the increase in the number of known solutions will be at the cost of easy-to-compute orbits. 
With 44 Taylor series order and 128 bit of precision, however, one can ``close'' one period well for each found solution. 
Remember that these are also the orders and precision used during the first stage of the search, but at the first stage they are used on a quarter period.

What are the previously found (old) equal-mass periodic orbits of the considered class? Three such solutions - $F_{14}(1,1,1)$, $F_{15}(1,1,1)$, $F_{27}(1,1,1)$ are found in \cite{Li:2019}. With the half-period approaches in \cite{Hristovi:2024, free_fall:2024} 35 such orbits with $T^*<70$  and 53 orbits with  $T^*<80$ were obtained respectively. The union of orbits from \cite{Hristovi:2024, free_fall:2024} is 66 orbits with $T^*<80$.
Note also that the search in \cite{Hristovi:2024} is in the Euler half-twist i.c. domain and uses the same search grid as in the present work.
All these previously found solutions are with $T^*<120$. They are rediscovered and contained in the new list.
We want to note that there are 9 other symmetric solutions in \cite{Li:2019}, which are not with a central, but with a reflectional symmetry. 
We do not consider this kind of orbits here. Description and figures of all 12 (3+9) symmetric solutions from \cite{Li:2019} can be seen in \cite{A. Gofen:2022}. 183 solutions with the other (reflectional) symmetry are found in \cite{free_fall:2024}.

\begin{figure}[h]
\centerline{\includegraphics[scale=0.52]{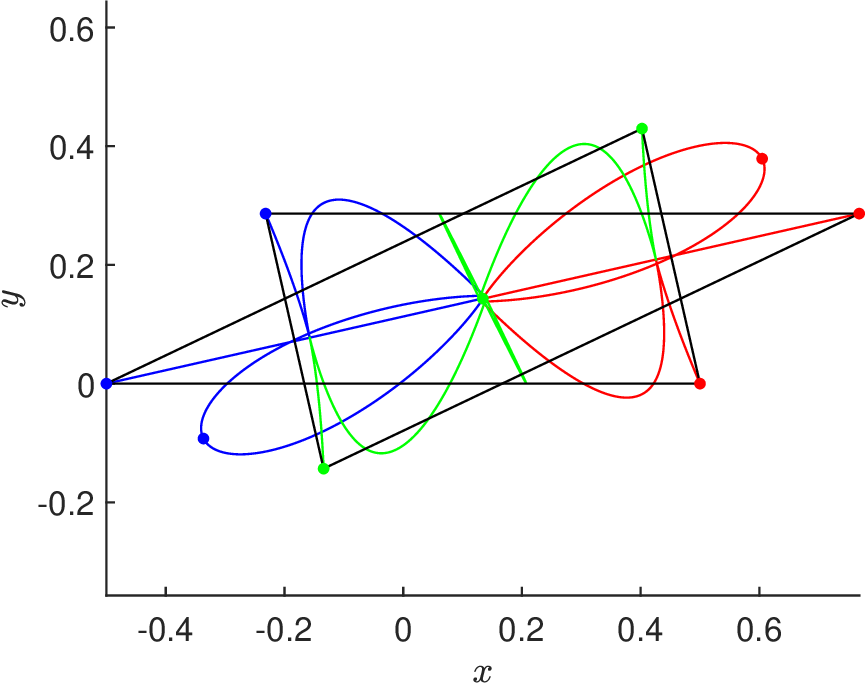}
\includegraphics[scale=0.52]{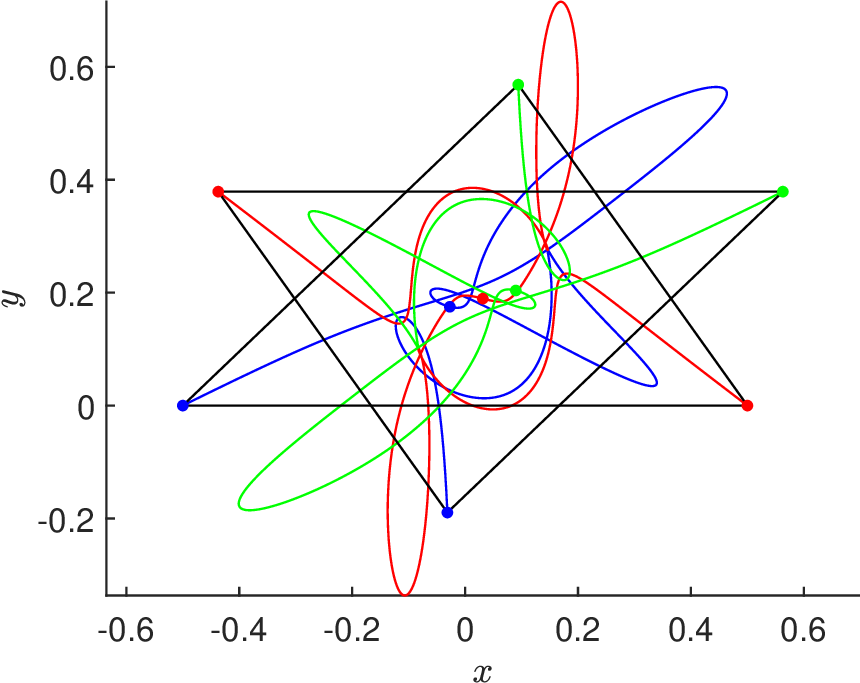}}
\vspace{0.52 cm}
\caption{{Real space plots for solution 7 (left) and 14 (right) from our list presented by free-fall i.c.}}
\label{fig:7and14}
\end{figure}

\begin{table}[h]
 \begin{tabular}{ p{7cm} p{7cm}}
 \hline
  $\hspace{3cm}v_x$ & $\hspace{3cm}v_y$\\
  $\hspace{3cm}T$ & $\hspace{3cm}T^*$\\
  \hline
  0.10848242901613985329364715023799301e-2 & 0.2901168948253795811356080552280542e0 \\
  0.1119580655126737921146136843283297e2 &  0.37722712752082117220726607574090248e2 \\
 \hline
  0.77023714561672049318242394420874335e0 & 0.40848239389471720919374725830660486e0 \\
  0.4388928697974290387170420610753859e3 & 0.45174968379100376402162574101966201e2 \\
 \hline
\end{tabular}
{\caption {35 correct digits for the Euler half-twist i.c. of solutions 7 and 14. }}
\label{tab:table3}
\end{table}

\begin{table}[h]
 \begin{tabular}{ p{7cm} p{7cm}}
 \hline
  $\hspace{3cm}p_x$ & $\hspace{3cm}p_y$\\
  $\hspace{3cm}T$ & $\hspace{3cm}T^*$\\
  \hline
  0.40230107184548177505098485089714165e0 & 0.42968944533765249707467569545897066e0 \\
  0.42753179670218930283536705782151218e1 & 0.37722712752082117220726607574090248e2 \\
 \hline
  0.94171067254084823214878198167535523e-1 & 0.56832432428452431039207305195301886e0 \\
  0.64831024875566300021372738476303985e1 & 0.45174968379100376402162574101966201e2 \\
 \hline
\end{tabular}
{\caption {35 correct digits for the free-fall i.c. of solutions 7 and 14.}}
\label{tab:table4}
\end{table}

With so many found orbits it is practically very difficult to inspect all their plots. 
It is very likely that some of the solutions have additional and interesting properties that could be a subject to further analysis. 
Anyway, we will give two plots of new solutions in order to show one property that is probably missed to have been clearly stated, albeit there are examples
of it  in \cite{free_fall:2024}. The property says that the interchanged masses in the ``stop'' triangles can be any two ones, equivalently, each of the three bodies can be in the middle for the Euler half-twist i.c. Real space plots of solutions 7 and 14 from our list are shown in Figure \ref{fig:7and14}.
The 35 correct digits data for these solutions is given in Tables 3 and 4. Let us assume blue, red and green colors against the shortest, middle length and longest side of the initial ``stop'' triangle. Then it is seen in Figure \ref{fig:F14} (right) and Figure \ref{fig:7and14} 
that all three color options for the middle body for the Euler half-twist i.c. are possible.

\begin{figure}[h]
 \begin{center}
 \includegraphics[scale=0.25]{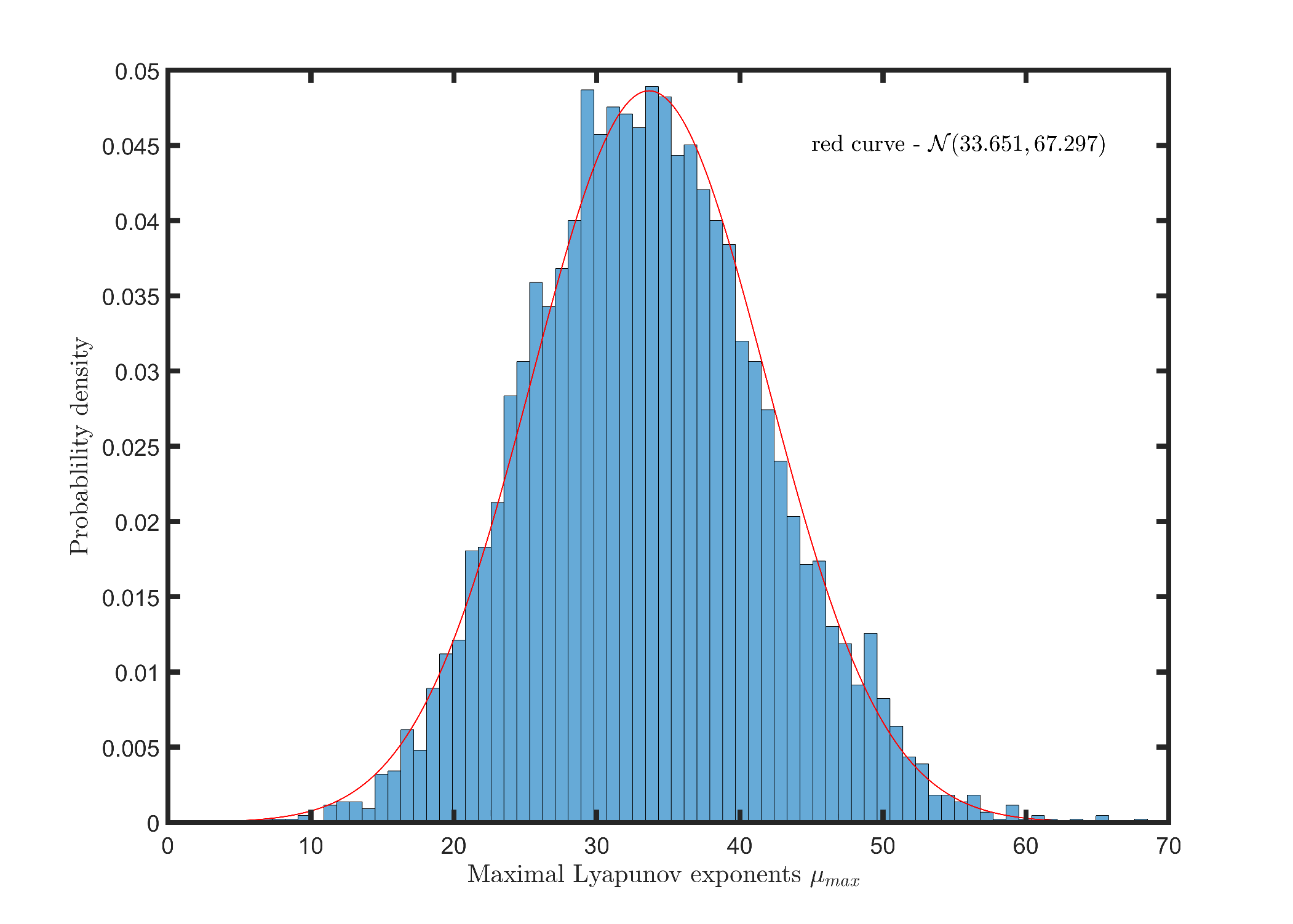}
 \caption{Distribution of the maximal Lyapunov exponents $\mu_{max}$}
 \label{Lyapexp}
 \end{center}
 \end{figure}

All 4860 orbits found are additionally investigated for linear stability by computing the corresponding monodromy matrices and finding their eigenvalues. 
Since the monodromy matrices are of the same (Jacobian) type as the matrices in Newton's method, they are computed in the same way as them - using the Taylor series method. To find the eigenvalues, we use the Multiprecision Computing Toolbox Advanpix \cite{Advanpix} for MATLAB\textregistered \cite{Matlab}.
Four of the eigenvalues determine the linear stability \cite{Roberts:2007}. These four eigenvalues, as well as the corresponding maximal Lyapunov exponents $\mu_{max}$ (all of them being scale-invariant quantities) can be found with 30 correct digits in \cite{Paper web-site:2025}.
All solutions are unstable: 4 of Loxodromic type, 198 of hyperbolic-elliptic type and the rest are of hyperbolic-hyperbolic type. 
The distribution of the maximal Lyapunov exponents $\mu_{max}$ is shown in Figure \ref{Lyapexp}. 
The minimal maximal Lyapunov exponent is $\approx6.512$, and the maximal is $\approx73.369$. The distribution fits well to a normal distribution with  mean 33.651 and variance 67.297. More details about linear stability computations are given in \cite{free_fall_stab:2024}.

\begin{figure}[h]
 \begin{center}
 \includegraphics[scale=0.38]{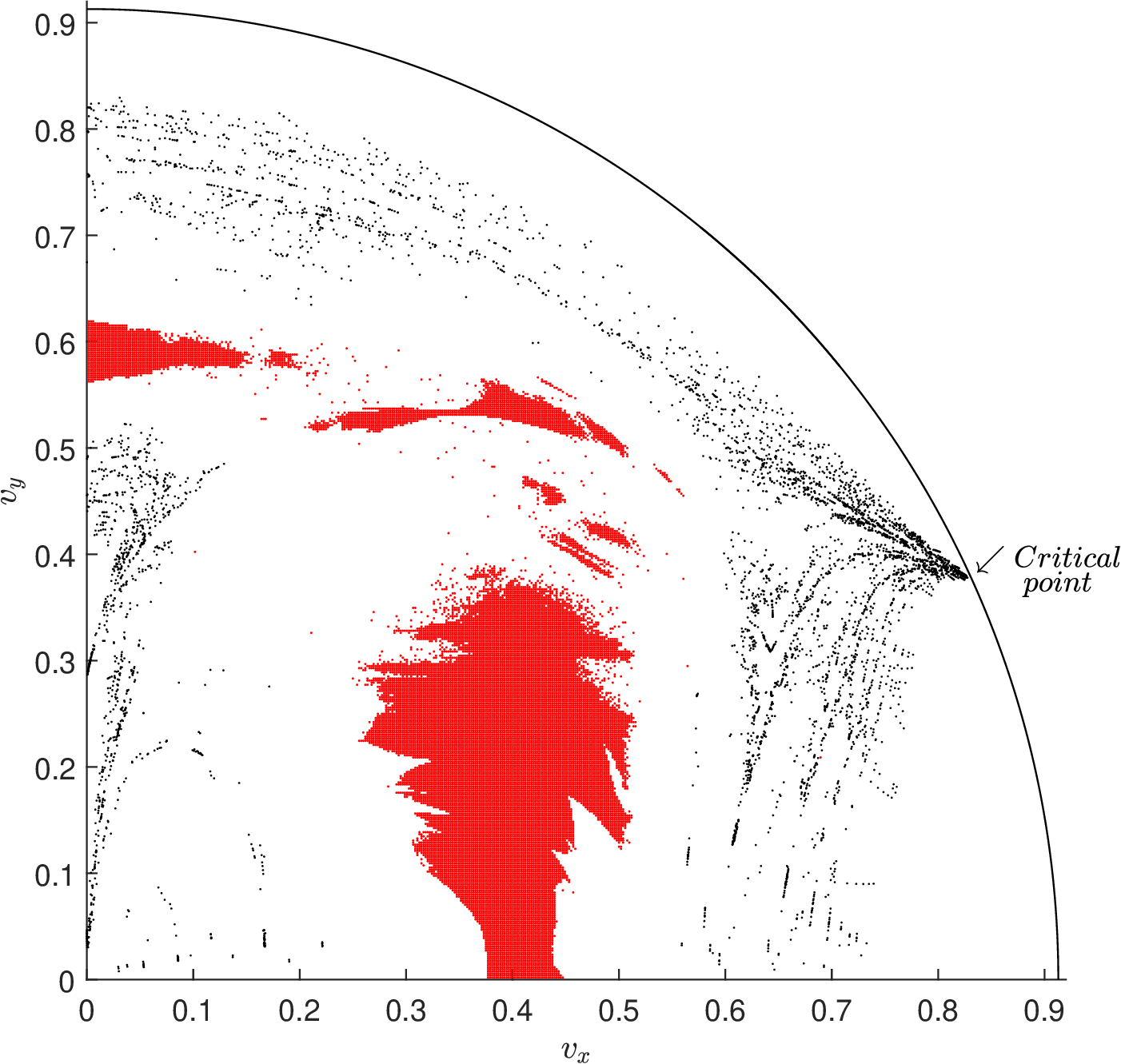}
 \caption{Distribution of the initial conditions in the Euler half-twist i.c. domain. In red is the stability region as defined in \cite{Martynova:2009}.
 The structure of the distribution is similar to those of the escape-times in \cite{Martynova:2009}.}
 \label{Distr_Euler}
 \end{center}
 \end{figure}
\begin{figure}[h]
 \begin{center}
 \includegraphics[scale=0.38]{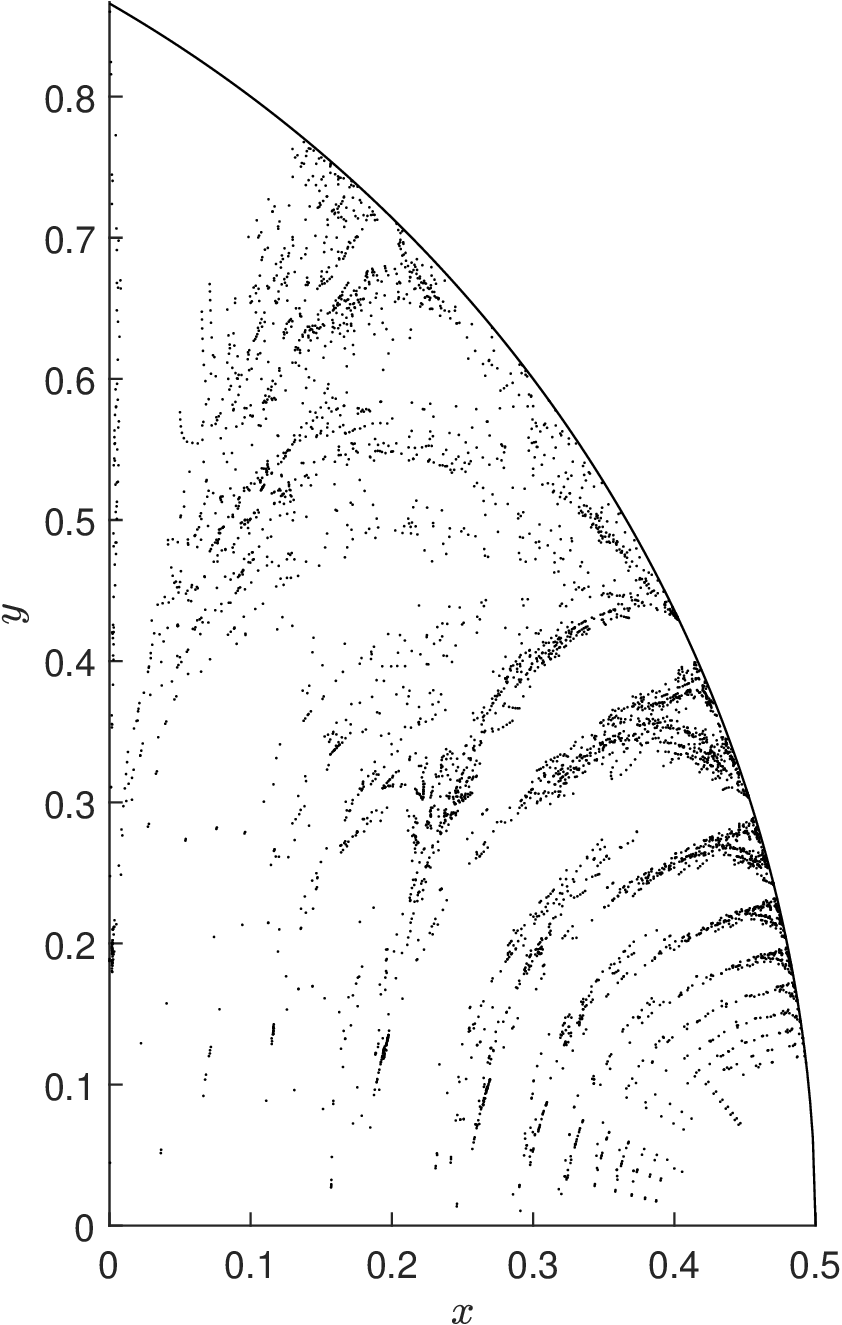}
 \caption{Distribution of the initial conditions in the free-fall i.c. domain. The structure of the distribution is similar to those of the escape-times in \cite{free_fall_stab:2024}.}
 \label{Distr_AA}
 \end{center}
 \end{figure}

The found initial conditions in the Euler half-twist i.c. domain are shown in Figure \ref{Distr_Euler} as black points. It can be observed that the black points are far from the red (stability) regions obtained in \cite{Martynova:2009}, which confirms the linear stability results from eigenvalues here, as well as the hypothesis stated in \cite{free_fall_stab:2024}, that the collisionless equal-mass free-fall periodic orbits are unstable. The red (stability) points, computed on a quadratic grid with a size of $2^{-9}$, are those points for which the scale-invariant time $t^*=10000$ is not sufficient for escape.  The used escape criterion is the same as in \cite{Martynova:2009}, i.e. the maximal distance between the bodies $r_{max}>5d$, where $d=3/|E|$ is the average triple system size. The red points in Figure \ref{Distr_Euler} are verified with our high precision ODE-solver (Taylor series method) by comparing the results for gradually increased precision and also with a reversibility test. The distribution of the initial conditions has a similar structure to that of the escape-times in \cite{Martynova:2009}, including the ``Critical point'' in \cite{Martynova:2009} on the circular boundary ($E=0$ boundary), in which  vicinity there is a cluster of points in the form of a spike. 

The  found initial conditions in the free-fall i.c. domain are shown in Figure \ref{Distr_AA}. It can be observed from this picture that the structure of their distribution is also similar to that of the escape-times in \cite{free_fall_stab:2024}, computed by the escape criterion in \cite{Standish:1971}.

One can find in \cite{Paper web-site:2025} the real space plots of the first 200 solutions in both the Euler half-twist i.c. and the free-fall i.c. domain. 
The syzygy sequences of all solutions computed by Tanikawa and Mikkola's syzygy counting method \cite{Tanikawa:2008,Tanikawa:2015} are also in \cite{Paper web-site:2025}. All intensive and high precision computations are performed in the HybriLIT platform, JINR, Dubna, Russia \cite{HybriLIT}.

\subsection{Discussion of the efficiency of the grid-search method in combination with Newton's method}
\label{ss:Discussion of the efficiency}

As explained in works  \cite{Li:2019, Li:2017, Li:2018} key for the successful simulation of the three-body problem is to overcome the sensitive dependence on the initial conditions, which is a property of chaotic systems.  To deal with this sensitivity, we use the proposed very successful ODE solver in \cite{Li:2019, Li:2017, Li:2018}  - the Taylor series method. But this decision is not sufficient for the efficient working of Newton's method applied with initial approximations obtained by the grid-search method. Replacing the solving of the periodicity equation with solving an equation defined on a part of the period, which also characterizes the periodic orbits, greatly affects the efficiency. Of course, each solution captured by solving an equation on a part of the period is tested after that for convergence with Newton's method for the standard periodicity equation, but then the initial conditions are already accurate enough and the convergence test is always passed.

Now we will give a simple Lyapunov exponents' argument, which  
explains the observed higher efficiency, when we solve an equation on a part of the period instead of solving the standard periodicity equation.
Let $p(t)$ be a given unstable periodic orbit with period $T$ and $\widetilde{p}(t)$ be an approximation of it.
Then for a small separation $d(t)=\|p(t)-\widetilde{p}(t)\|_{2}$ 
between the adjacent trajectories the exponential law of divergence is satisfied: 
$$d(t)\approx d(0) e^{\lambda t}, \quad \lambda - \text{the Lyapunov exponent}$$
Here $\lambda T \approx \mu_{max}$, where $\mu_{max}$ are the computed maximal Lyapunov exponents from the
eigenvalues of the monodromy matrices \cite{free_fall_stab:2024}.
If we take for example $d(0)\approx 10^{-4}$ and take $e^{\lambda T}\approx10^8$,
then $d(T/2) \approx 1$, but $d(T) \approx 10^4.$
In the first case we can expect convergence of Newton's method. In the second case the above approximate exponential low is in fact not valid.  
We can expect that in this case the trajectories are already so mixed at $t=T$ that Newton's method
does not work (regardless of the fact that we have computed the trajectories accurately). Note that to obtain a similar result for the example case $e^{\lambda T}\approx10^8$ by solving  the standard periodicity equation, we have to take $d(0)\approx 10^{-8}$, which means to use such a fine search grid that the high precision computations are much more time consuming.

The effect of dividing the period by three and by four is even stronger.
Instead of the square root of $e^{\lambda T}$ we take its third and fourth root respectively.
For the considered initial distance $d(0)\approx 10^{-4}$ Newton's method should work well up to even greater Lyapunov exponents: $e^{\lambda T}\approx10^{12}$
and $e^{\lambda T}\approx10^{16}$ respectively. We can say that the smaller the  considered part of the period, the better the results.

To get a better feel for how things are, let us assume the simplification that $T$ is exact in the approximation $(v_x,v_y,T)$ 
and the upper limit $T^*_{max}$ for the scale-invariant periods is fixed. Then around the points $(v_x, v_y)$ (the exact solutions) in the Euler half-twist i.c. domain there are certain $2D$ vicinities (regions) for which Newton's method converges to the corresponding solution. The vicinities are smaller for larger Lyapunov exponents. Of course, there are also regions  for which Newton's method does not converge.  
By replacing the periodicity equation with that on half, third, and quarter period, the regions of convergence gradually expand, and those of non-convergence decrease. Which solutions will be captured depends on: (1) what is the considered class of orbits, (2) what $T^*_{max}$ we choose, (3) what grid we choose, (4) what equation we solve (on the entire period or on a part of it), (5) what exactly the criterion is for the candidates for correction with Newton's method. 
The number of the found orbits and the distribution of the maximal Lyapunov exponents change accordingly to the mentioned above assumptions. 
For fixed $T^*_{max}$ the mean value of the maximal Lyapunov exponents is expected to increase when the number of the found orbits increases (the efficiency of the search increases). 

Let us compare the maximal Lyapunov exponents for the old 66 periodic orbits from the considered class with $T^* <80$ found in \cite{Hristovi:2024, free_fall:2024} and the newly discovered ones in this work. Actually, the large number of found orbits here (4860) are obtained not only because of the used quarter-period approach, but also because longer periods are considered. Considering a narrower class of orbits allows longer periods to be considered for obtaining a reasonable (for computations and analysis) number  of solutions. In fact, only 312 orbits from 4860 have $T^*<80$, which is about 5 times more than the union of 66 orbits from \cite{Hristovi:2024, free_fall:2024}. Note, however, that for the 66 orbits the mean of maximal Lyapunov exponents is $ \approx 18.5$, with minimal and maximal values $ \approx 6.512$ and $ \approx 26.765$ respectively, but the added 246 orbits from this search have a mean of maximal Lyapunov exponents 
$\approx 34.326  (\approx 2 \times 18.5)$, with minimal and maximal values $ \approx 11.891$ and $ \approx 51.547$ respectively.
Another interesting observation that supports the efficiency increase for the quarter-period approach is the following one. 
If we allow in the case when Newton's sequence of approximate solutions initially ``jumps'' somewhere in the search domain
to keep trying for convergence and it finally converges, then such new (but randomly found) solutions are obtained much more often than before. 
In fact, the total number of newly discovered solutions is much over 5000, but we removed the randomly obtained solutions with $T^*>120$. 

For completeness we will recall the remaining results for the Lyapunov exponents. For the whole class of free-fall orbits with $T^*<80$ in \cite{free_fall:2024,free_fall_stab:2024} the distribution of the maximal Lyapunov exponents of the found solutions fits well to a normal distribution with mean $20.407$  and variance $13.831$. Similarly, for the Euler half-twist initial conditions' orbits with $T^*<70$  in \cite{Hristovi:2024} the distribution fits well to a normal one with mean $17.571$ and variance $16.393$. For the narrower class of free-fall orbits with central symmetry in this work and $T^*<120$ the distribution fits well to a normal one with mean $33.651$  and variance $67.297$ (see Figure \ref{Lyapexp}). 
Note that $e^{20.407} \approx 7.3*10^8, e^{17.571} \approx 4.3*10^7, e^{33.651} \approx 4.1*10^{14}$, which supports our simple Lyapunov exponents' explanation for the increase of efficiency. These results show us that generally we have to be very careful about the conclusions we draw from the obtained maximal Lyapunov exponents (and of course, all other quantities characterizing the obtained solutions), because they can be a result of an imperfect search on a grid.

From the computational experience until now, we can say that for the classes of orbits and relatively short periods considered so far, the 44 order of Taylor series and 128 bit of precision used in the first (scanning) stage seem to be a sufficiently good choice, because all (or almost all) orbits can ``close'' a full period well with this order and precision, while only half, third or a quarter of the period are considered in the scanning stage and also the precision in the subsequent stages is increased. Still, we expect the need of order and precision for the scanning stage to be increased when longer periods are considered. On the other hand, the considered search grids for the previous half-period approaches \cite{Hristovi:2024, free_fall:2024, free_fall_stab:2024}, selected with the goal of solving the problem in a foreseeable time, seem to not be dense enough for the precision chosen. The results in \cite{Hristovi:2024, free_fall:2024, free_fall_stab:2024}, although much better than the previous ones, seem to be still not very close to exhaustive searches (if they are at all possible) for periodic orbits with the considered limited periods.

We would also like to say few words in relation to the recent artificial intelligence (AI) paper \cite{Liao:2022}.
The proposed efficient approaches in the sequence of works - this one and the previous ones\cite{Hristovi:2024, free_fall:2024, free_fall_stab:2024, Grid:2023 }, address the efficiency of the roadmap's first stage in paper \cite{Liao:2022}, before applying AI. We think that our works and \cite{Liao:2022} complement each other well.

\section{Concluding remarks}
\label{s:Conclusions}
$_{}$

1) An  efficient quarter-period approach for a specialized numerical search for equal-mass collisionless three-body periodic free-fall orbits with central symmetry is proposed. The conducted search significantly enlarged the number of the known orbits from the class.
The search produced 4860 orbits with $T^*<120$ presented as a high precision (100 correct digits) database.
The distribution of the initial conditions in the search domains shows a similar structure to those of the previously computed escape-times.

2) The linear stability of the orbits is also numerically investigated. All found orbits are unstable. This result confirms the hypothesis that equal-mass collisionless three-body periodic free-fall orbits are unstable. The eigenvalues that determine the linear stability and the corresponding maximal Lyapunov exponents are presented as a high precision (30 correct digits) database.

3) Using a simple Lyapunov exponents' argument and analyzing the maximal Lyapunov exponents of the obtained orbits, it is shown that solving equations only on a part of the period instead of solving the periodicity equation on the entire period has a very strong effect on the efficiency of Newton's method applied to initial approximations obtained by the grid-search method. The smaller the part of the period, the better the results. This means that the discovery and proof of characteristic properties (usually symmetry properties) of the periodic orbits from a given class, that are expressed by equations defined on a part of the period, is an important mathematical task. Solutions to this task could help a lot in the increase of efficiency of future numerical searches.

\section*{Acknowledgement}
The authors acknowledge the access to the computational resources of the HybriLIT platform, JINR, Dubna, Russia.
I.H. thanks Richard Montgomery for answers to theoretical questions and advices.
The work of I.H. is supported by the European Union-NextGenerationEU,
through the National Recovery and Resilience Plan of the Republic of Bulgaria, project number BG-RRP-2.004-0008-C01.

\end{document}